\documentclass[%
aip,
amsmath,amssymb,
reprint,%
]{revtex4-2}

\usepackage{graphicx}
\usepackage{dcolumn}
\usepackage{bm}

\usepackage[utf8]{inputenc}
\usepackage[T1]{fontenc}
\usepackage{mathptmx}
\usepackage{etoolbox}

\usepackage[dvipsnames]{xcolor}
\usepackage[normalem]{ulem}

\makeatletter
\def\@email#1#2{%
	\endgroup
	\patchcmd{\titleblock@produce}
	{\frontmatter@RRAPformat}
	{\frontmatter@RRAPformat{\produce@RRAP{*#1\href{mailto:#2}{#2}}}\frontmatter@RRAPformat}
	{}{}
}%

\makeatother
\renewcommand{\selectlanguage}[1]{}
\begin{document}
	
	\preprint{AIP/123-QED}
	
	\title[Compact structures for single-beam magneto-optical trapping of ytterbium]{Compact structures for single-beam magneto-optical trapping of ytterbium}
	\author{J. Pick}
	\email[]{julian.pick@dlr.de}
	\affiliation{Deutsches Zentrum für Luft- und Raumfahrt e.V., Institut für Satellitengeodäsie und Inertialsensorik, Callinstraße 30b, 30167 Hannover, Germany}
	\affiliation{Leibniz Universität Hannover, Institut für Quantenoptik, Welfengarten 1, 30167 Hannover, Germany}
	\author{R. Schwarz}
	\author{J. Kruse}
	\affiliation{Deutsches Zentrum für Luft- und Raumfahrt e.V., Institut für Satellitengeodäsie und Inertialsensorik, Callinstraße 30b, 30167 Hannover, Germany}
	\author{C. Lisdat}
	\affiliation{Physikalisch-Technische Bundesanstalt, Bundesallee 100, 38116 Braunschweig, Germany}
	\author{C. Klempt}
	\affiliation{Deutsches Zentrum für Luft- und Raumfahrt e.V., Institut für Satellitengeodäsie und Inertialsensorik, Callinstraße 30b, 30167 Hannover, Germany}
	\affiliation{Leibniz Universität Hannover, Institut für Quantenoptik, Welfengarten 1, 30167 Hannover, Germany}

	\date{\today}
	
	\begin{abstract}
	Today's best optical lattice clocks are based on the spectroscopy of trapped alkaline-earth-like atoms such as ytterbium and strontium atoms. The development towards mobile or even space-borne clocks necessitates concepts for the compact laser-cooling and trapping of these atoms with reduced laser requirements. Here we present two compact and robust achromatic mirror structures for single-beam magneto-optical trapping of alkaline-earth-like atoms using two widely separated optical cooling frequencies. We have compared the trapping and cooling performance of a monolithic aluminium structure that generates a conventional trap geometry to a quasi-planar platform based on a periodic mirror structure for different isotopes of Yb. Compared to prior work with strontium in non-conventional traps, where only bosons were trapped on a narrow line transition, we demonstrate two-stage cooling and trapping of a fermionic alkaline-earth-like isotope in a single-beam quasi-planar structure. 
	\end{abstract}
	
	\maketitle

	\section{\label{sec:introduction}Introduction}
    Quantum sensors such as optical clocks undergo a continuous transformation from laboratory-based systems to field-deployable instruments and even reaching out into space \cite{bongs_development_2015,schkolnik_optical_2023,tino_sage_2019}. Along with the downsizing of key components, improving the systems' robustness and reliability is a crucial requirement. 
    Due to the presence of an ultra-narrow transition in the optical wavelength range, alkaline-earth-like atoms are used in optical lattice clocks (OLC) \cite{takamoto_optical_2005, mcgrew_atomic_2018, bothwell_resolving_2022, schwarz_long_2020}. Recent efforts in reduction of size, weight and power consumption (SWaP) of OLC key components enabled the deployment of transportable systems \cite{ohmae_transportable_2021, koller_transportable_2017, fasano_transportable_2021, origlia_optical_2018}. 
    Quantum sensors based on ultra-cold atoms typically rely on the trapping and cooling of atoms in a magneto-optical trap (MOT). While the well known six-beam configuration \cite{raab_trapping_1987} requires a rather extensive optical setup with up to six individual optical beams, simplification can be achieved by using in-vacuum optics. Several compact structures have been demonstrated which enable three-dimensional trapping with a single incident beam only. Assisted by the structure itself, several secondary beams are formed from the incident beam by either reflective or diffractive optical elements. Prominent compact MOT geometries using reflective optics are the so-called ``pyramid MOT'' \cite{lee_singlebeam_1996,bowden_pyramid_2019} and the ``tetrahedral MOT'' \cite{vangeleyn_singlelaser_2009}. A higher level of compactness can be achieved by operating MOTs with planar diffraction gratings \cite{vangeleyn_laser_2010, barker_grating_2023a, bondza_twocolor_2022,sitaram_confinement_2020,burrow_optimal_2023a}, often referred to as ``gMOTs''. These structures offer unrestricted radial access to manipulate the trapped atoms with additional optical beams. Nevertheless, when two-color operation is required e.g. for cooling on two widely separated atomic transitions, they come at the cost of substantial chromaticity. This leads to a significant spatial mismatch of the corresponding trapping volumina and causes difficulties in trapping. Additionally, the fabrication of a gMOT is comparably expensive due to the photolithography process. In contrast, metallic reflective structures can be manufactured by conventional mechanical machining.
    While reflective elements offer the advantage of achromaticity, the recently demonstrated Fresnel reflector \cite{bondza_achromatic_2024,bondza_atomfalle_2022} combines this advantage with being planar.
    
    Of the aforementioned optics, only the pyramid reflector with its secondary beams emerging perpendicular to the incident beam maintains the conventional MOT-beam geometry leading to the same trapping dynamics as the six-beam configuration. For the Fresnel reflector, the secondary beams form an oblique angle with the magnetic field, resulting in the presence of additional polarizations seen by the atoms which complicates the cooling and trapping dynamics. This can be a reason why, so far, neither of the clock-relevant fermions, strontium or ytterbium, were trapped on a narrow-line cooling transition in such a non-conventional geometry. For strontium, this was only demonstrated with bosonic isotopes \cite{elgee_grating_2022, bondza_twocolor_2022, bondza_achromatic_2024}, while for ytterbium, so far no MOT operation in any single-beam configuration was reported.
    
    In this article, we demonstrate compact cooling and trapping of ytterbium atoms in two different single-beam MOT geometries. We compare the results for a pyramid reflector and a Fresnel reflector when applied for the trapping of a variety of bosonic and fermionic Yb isotopes.
    Our results can be applied for designing a new generation of highly compact ytterbium-based optical lattice clocks.

	\section{Reflector Geometries}
	\begin{figure*}
	\includegraphics[width = 0.9\textwidth]{"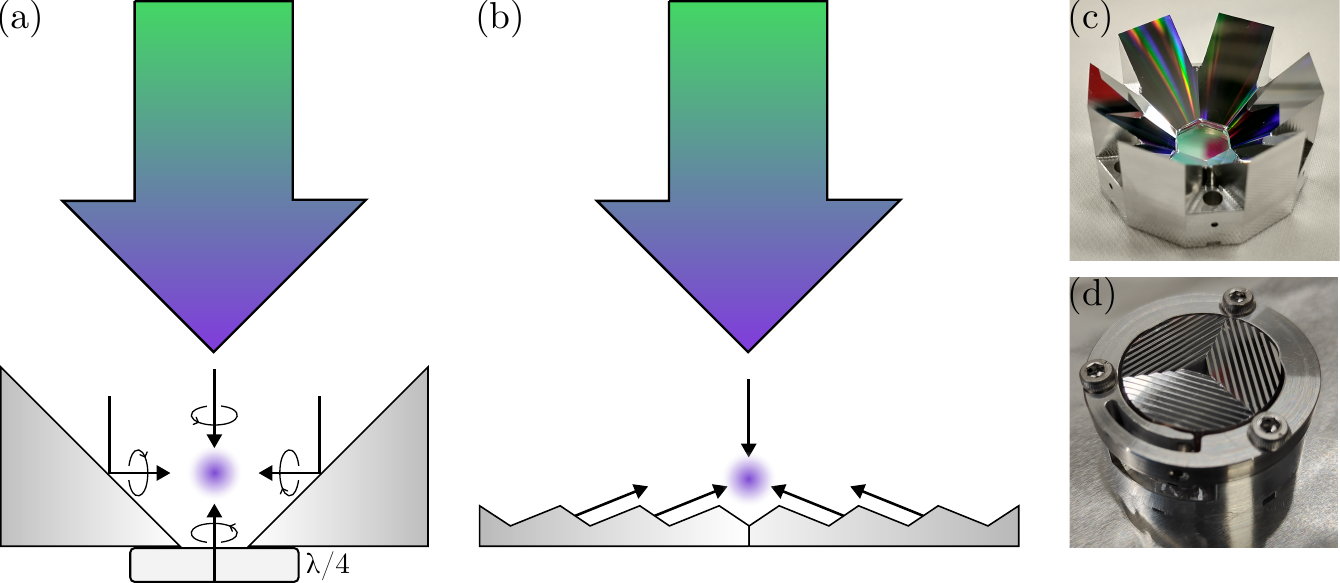"}
	\caption{\label{fig:mot_geometries}Tested single-beam MOT geometries. (a) Pyramid reflector. A single incident beam is reflected by six angled surfaces to generate the radial MOT beams. A bichromatic waveplate with a highly-reflective coating on the back side provides the axial reflected beam with the required polarization handedness. (b) Fresnel reflector. It consists of three segments with a periodic mirror structure. The reflective surfaces are angled close to the ideal tetrahedral configuration, ensuring intensity balance between the incident beam and the reflected secondary beams. (c) Photo of the pyramid reflector. (d) Photo of the Fresnel reflector assembled inside its flexure mount.}
	\end{figure*}

	Although the pyramid and the Fresnel reflector both generate the secondary beams by reflection of a single incident beam, they are designed very disparately. Furthermore, the corresponding cooling and trapping dynamics differ fundamentally due to the different orientations of the secondary beams with respect to the magnetic field axes. Both structures are described in detail in the following subsections.
	
	\subsection{The pyramid reflector}
	The pyramid reflector is depicted in Fig.~\ref{fig:mot_geometries}a and \ref{fig:mot_geometries}c. Its hexagonal symmetry is inspired from Ref. \cite{bowden_pyramid_2019} and gives three radial axes for optical access and atom loading. However, instead of assembling individual glass prisms, we use a monolithic design. This increases robustness and guarantees optimal alignment of the mirror surfaces. The reflector structure was milled from a single block of aluminium and the angled surfaces were successively polished in order to reflect the incident beam to form the radial MOT beams. No coating is added to the polished surfaces, since aluminium is highly reflective in the visible spectrum, particularly for both relevant wavelengths of $399\,$nm and $556\,$nm. 
	The reflector has an outer diameter of $34\,$mm and includes six screw holes for direct mounting into a vacuum chamber. For this monolithic design, the manufacturing of the pyramid apex would pose a significant technical challenge and induce a risk for the effective MOT operation. We overcome this challenge by replacing the apex with a bichromatic $\lambda/4$-waveplate that can be inserted into the aluminium structure from below. This waveplate has an anti-reflective coating on the front side and a highly-reflective coating on the back side, so that the reflected beam passes the waveplate twice and obtains the required polarization handedness. In contrast to the retroreflector prism in the design from Ref.\cite{bowden_pyramid_2019}, this waveplate is the only part of the reflector that is not achromatic, preventing multi-species operation. However, if operation at a different wavelength is desired, it can be easily replaced without any changes to the reflector design.
	
	\subsection{The Fresnel reflector}
	The Fresnel reflector is described in detail in Ref. \cite{bondza_achromatic_2024} and is depicted in Fig.~\ref{fig:mot_geometries}b and \ref{fig:mot_geometries}d. It consists of three segments, which possess a periodic mirror structure, so that a single incident beam is split and reflected towards the trap center. The reflective surfaces are angled at 34°, close to the ideal tetrahedral configuration. In fact, the Fresnel reflector is a quasi-planar version of the tetrahedral mirror MOT \cite{vangeleyn_singlelaser_2009}, yielding the same beam geometry and trapping dynamics. This quasi-planar design gives a higher level of compactness and unrestricted radial optical access.
	The three segments were machined from oxygen-free copper, and were successively coated with aluminium. The reflector has a diameter of $25.4\,$mm.
	
	While the pyramid reflector generates a beam geometry that is equivalent to a conventional six-beam MOT, this is not the case for the Fresnel reflector. The secondary beams form an oblique angle with the magnetic field axes, so that they can couple to all ($\sigma^{+}$, $\pi$, $\sigma^{-}$) transitions, even if the beams had purely circular polarizations. The resulting implications on the MOT operation are extensively covered in the literature \cite{barker_grating_2023a, bondza_twocolor_2022, vangeleyn_singlelaser_2009, vangeleyn_laser_2010, vangeleyn_atom_2011}. A major difference to the conventional MOT geometry is that the effective trap depth is reduced and that the cooling and trapping dynamics become increasingly complicated for isotopes with a rich hyperfine structure. For alkaline-earth-like elements, these are the fermionic isotopes, which are commonly used in OLCs. This challenge presents a major motivation for the experimental investigation of two-stage cooling of ytterbium in such a non-conventional MOT geometry.

	\section{First-stage cooling and trapping}
	The atom source used for loading the MOTs is a copy of the one described in Ref. \cite{wodey_robust_2021}. Fast atoms emerge from an oven operated around $480\,^\circ$C and are subsequently slowed by a permanent-magnet Zeeman slower operating on the broad $^1S_0 \rightarrow \,^{1}P_1$ transition at $399\,$nm. The Zeeman slower beam has a $1/e^2$ diameter of $4\,$mm, a power of $100\,$mW and a detuning of $-660\,$MHz from resonance at rest. The slowed atoms escape from the slower at a velocity of around $20\,$m/s. They are subsequently redirected and recollimated by a 2D-MOT, loaded under a 30° angle, separating the Zeeman slower beam from the MOT volume. The 2D-MOT beams have a cylindrical shape with $1\,$cm $\times$ $4\,$cm diameter, a power of $45\,$mW and a detuning of $-20\,$MHz. The atomic flux emerging from the 2D-MOT is $7\cdot 10^9\,\mathrm{atoms}/\mathrm{s}$ for $^{174}$Yb. For the different isotopes, the relative flux agrees with the respective natural abundance, except for $^{173}$Yb, where the flux is less than $40\,\%$ of the expected value.
	
	The Fresnel reflector has a central hole that can be used for atom loading from below. However, for comparability of the two reflectors, we decided to load both from the radial direction. The pyramid reflector was oriented such that the atomic beam passes between two reflective surfaces.
	The reflectors are both illuminated with a bichromatic beam at $399\,$nm and $556\,$nm. The transverse intensity distribution of this beam has a major impact on the trapping dynamics for the Fresnel reflector \cite{barker_grating_2023a} and also defines the intensity ratio between the radial and the axial beams for the pyramid reflector. A desirable beam shape is a flat-top beam, where the uniform intensity distribution results in the same intensities of the incident and the secondary beams. However, shaping a bichromatic beam to a large flat-top profile is a significant technical challenge and therefore we instead used a Gaussian beam collimated at a diameter of $45\,$mm using achromatic lenses. Because this diameter is much larger than the reflectors' diameters, the intensity distribution in the center is sufficiently flat.
	
	The required quadrupole magnetic field is generated by a pair of anti-Helmholtz coils, wound on water-cooled copper mounts. The typical magnetic field gradient used for the pyramid reflector is $3.5\,\mathrm{mT}/\mathrm{cm}$ in axial direction. For the Fresnel reflector, a larger gradient is required, typically $5.5\,\mathrm{mT}/\mathrm{cm}$.
	The vacuum pressure was below $5\cdot 10^{-10}\,$mBar during all experiments.
	
	\begin{figure}
		\centering
		\includegraphics[width = 0.45\textwidth]{"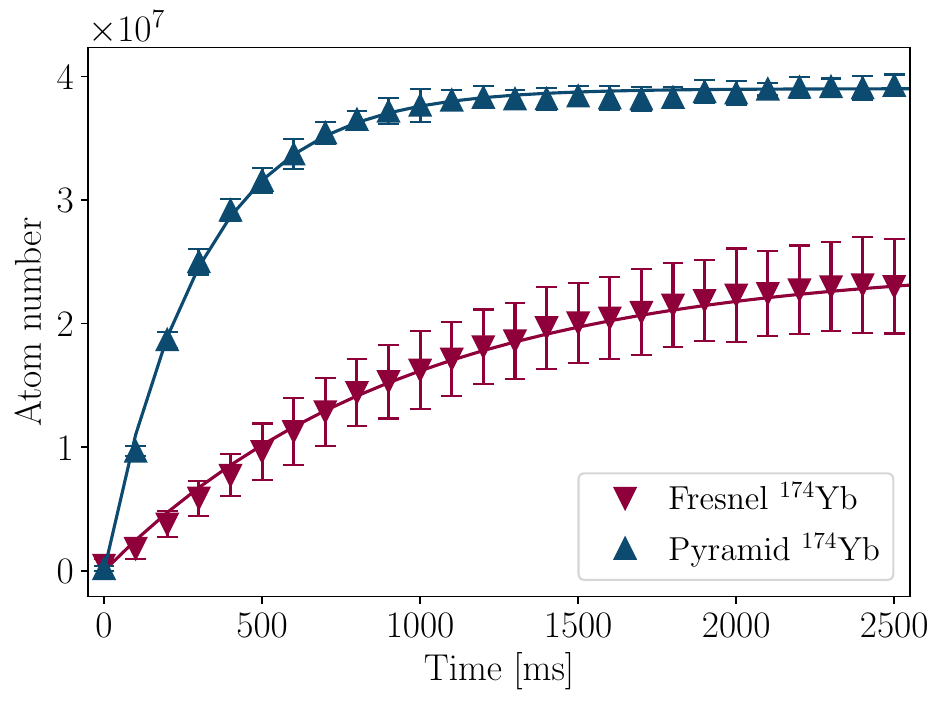"}
		\caption{\label{fig:399_loading_curves}First stage MOT loading curves for $^{174}$Yb with both reflectors at an optical power of $30\,$mW. Solid lines are exponential fits that yield a loading rate of $1.3\cdot10^8\,\mathrm{atoms}/\mathrm{s}$ for the pyramid and $2.6\cdot10^7\,\mathrm{atoms}/\mathrm{s}$ for the Fresnel reflector.}
	\end{figure}
	
	\begin{figure}
		\centering
		\includegraphics[width = 0.45\textwidth]{"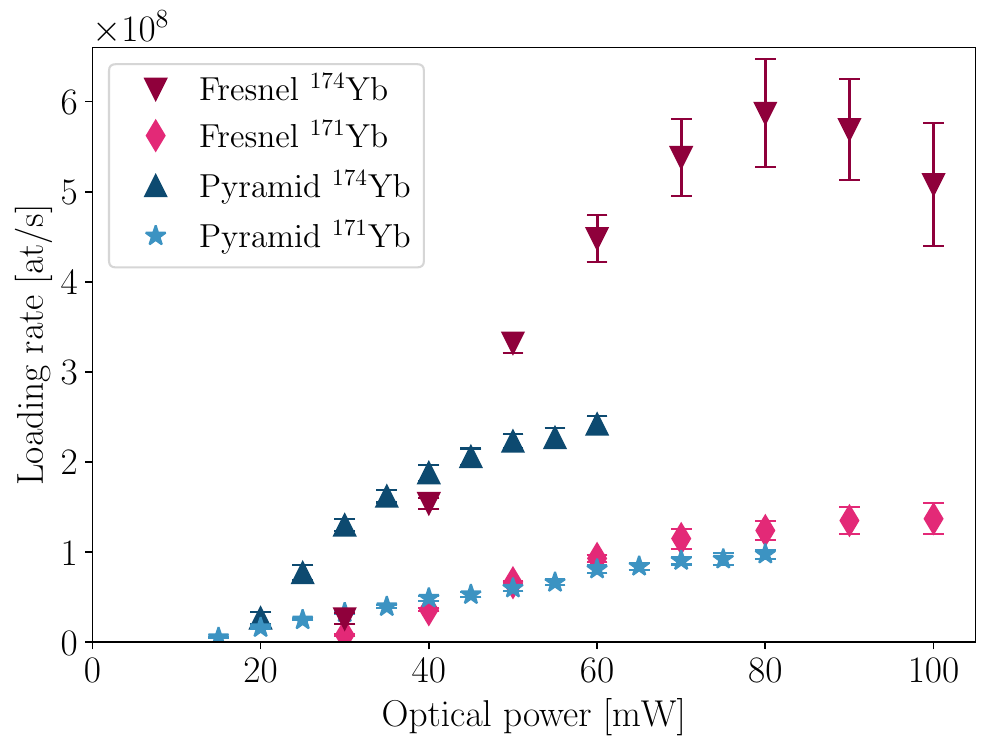"}
		\caption{\label{fig:399_loading_vs_power}First stage MOT loading rate as a function of optical power in the incident MOT beam  at $399\,$nm for the two reflectors and the isotopes $^{174}$Yb and $^{171}$Yb.}
	\end{figure}
	The optimal detunings for the first MOT stage operating at $399\,$nm differ for both the reflector type and the chosen isotope. For the most abundant isotope $^{174}$Yb, the optimal detuning was found to be $-34\,$MHz ($-24\,$MHz) for the pyramid (Fresnel) reflector. For the fermionic isotope $^{171}$Yb, it is $-22\,$MHz ($-20\,$MHz). Figure~\ref{fig:399_loading_curves} shows exemplary loading curves for both reflectors at the same optical power of $30\,$mW. Error bars are statistical and represent one standard deviation. From exponential fits the loading rates can be determined. We measured loading curves for different optical powers for both reflectors and isotopes and the corresponding loading rates are shown in Fig.~\ref{fig:399_loading_vs_power}. It becomes apparent that the pyramid outperforms the Fresnel reflector at a power below $40\,$mW. This is because of the higher number of reflected beams, which result in a larger intensity seen by the atoms for the same optical power. This means that the pyramid more efficiently recycles the incident power, which can become significant for transportable experiments, where the available power might be limited. However, at higher powers, the loading rate for the pyramid stagnates, while it further increases for the Fresnel reflector. This is due to the limited flux of atoms reaching the trapping region of the pyramid. The pyramid reflector truncates the atomic beam that is loaded from the radial direction, while this is not the case for the Fresnel reflector loaded radially. Although this is a technical issue that can be mitigated by an optimization of the atom source, this highlights an advantage of the Fresnel reflector, which is the larger radial access. This is not only beneficial for atom loading, but also for accessing the trapped atoms with all the laser beams that are further required for the operation of an OLC.
	The loading rate of $^{174}$Yb for the Fresnel reflector shows a maximum at $80\,$mW and decreases for higher powers. This behaviour agrees with the theoretical considerations in Ref. \cite{barker_grating_2023a}, where an increased coupling to the anti-trapping transitions is predicted to lead to a decrease in trapping forces. For the fermionic isotope, the improved performance of the Fresnel reflector at high optical power is less clear. This is connected to the more complicated trapping dynamics emerging from the coupling of the secondary beams to all ($\sigma^{+}$, $\pi$, $\sigma^{-}$) transitions, which is not the case for the pyramid. For all four combinations of reflector and isotope, it is possible to obtain a loading rate above $10^8\,\mathrm{atoms}/\mathrm{s}$, making both realizations promising source concepts for future compact optical clocks.

	\begin{figure}
		\centering
		\includegraphics[width = 0.45\textwidth]{"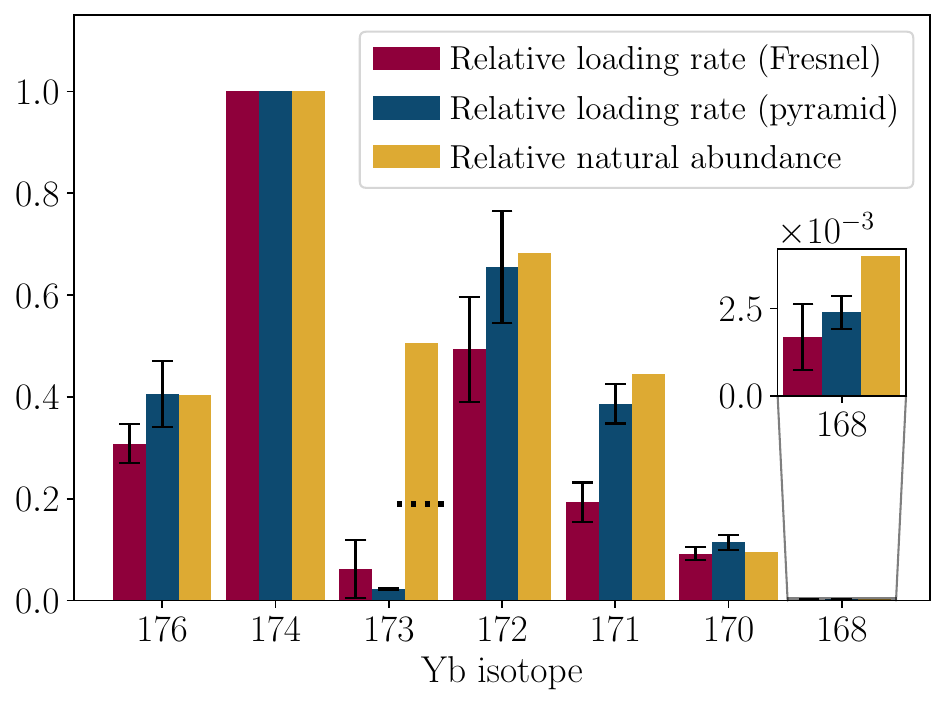"}
		\caption{\label{fig:399_loading_all_isotopes}Relative first-stage MOT loading rates for both reflectors and relative natural abundances of all stable isotopes of Yb. All values are given relative to the respective value of $^{174}$Yb. The black dotted line indicates the reduced atomic flux of $^{173}$Yb emerging from the 2D-MOT.}
	\end{figure}

	We load all stable isotopes into the first MOT stage with both reflectors. A comparison of the relative loading rates together with the relative natural abundances is shown in Fig.~\ref{fig:399_loading_all_isotopes}. Here, all values are normalized to the respective value of $^{174}$Yb. For each reflector, the loading rates of all isotopes are measured using the same magnetic field gradient and saturation parameter, $s=0.3$ for the pyramid and $s=0.2$ for the Fresnel reflector, while the detunings are optimized individually. Additionally, for the fermionic isotopes $^{171}$Yb and $^{173}$Yb, the Zeeman slower detuning and power are optimized, because here the optimal values differ from the bosonic isotopes.
	
	It becomes apparent that for the pyramid, the loading rates of the bosonic isotopes agree well with the natural abundances. For the Fresnel reflector, the deviations from the expected values are slightly larger. A major difference between the two reflectors is visible for $^{171}$Yb. While the relative loading rate for the pyramid is close to the relative natural abundance, it is clearly lower for the Fresnel reflector. This again shows that the MOT operation with a non-conventional beam geometry can be less efficient for fermionic isotopes. With both reflectors, the loading rate of $^{173}$Yb is significantly lower than what could be expected from the natural abundance and from the flux emerging from the 2D-MOT. This effect has also been observed in other Yb experiments \cite{dorscher_creation_2013, loftus_simultaneous_2001, honda_magnetooptical_1999} and is explained by optical pumping in the $^1S_0 (F=5/2)$ ground state and the proximity of the $F=5/2 \rightarrow F'=3/2$ transition to the $F=5/2 \rightarrow F'=7/2$ cooling transition. The successful operation of both MOT geometries with all relevant Yb isotopes demonstrates the broad applicability of the presented compact trap setups.
	
	\section{Second-stage cooling and trapping}
	The second MOT stage operates on the $^1S_0 \rightarrow \,^{3}P_1$ transition at $556\,$nm. After loading into the first MOT stage, the $556\,$nm light is switched on and after a transfer time of $20\,$ms, the $399\,$nm light is switched off. No frequency broadening of the $556\,$nm light is required for efficient transfer. We load the most commonly used isotopes $^{174}$Yb and $^{171}$Yb into the second MOT stage with both reflectors and although the first-stage trapping of $^{173}$Yb was not very efficient, we successfully transfer it to the second MOT stage with the pyramid reflector. We investigate the transfer efficiencies and lifetimes by loading atoms into the second MOT stage, holding them for a variable time and subsequently transferring them back. By comparing the initial and final population in the first MOT stage, we can measure the fraction of atoms in the second MOT stage as a function of hold time. From this, we can calculate the initial transfer efficiency and the lifetime in the second MOT stage.
	
	We also measure the temperature of the atomic cloud after second-stage cooling. To this end, we switch off the MOT light and the magnetic field gradient and let the cloud expand ballistically. After a variable expansion time, the cloud is illuminated with resonant light at $399\,$nm for an exposure time of $100\,$µs and a camera fluorescence image is taken. The Gaussian width of the cloud is fitted to the image. The cloud width as a function of expansion time then yields the temperature.
	
		\subsection{Pyramid reflector results}
		\begin{figure}
			\includegraphics[width = 0.45\textwidth]{"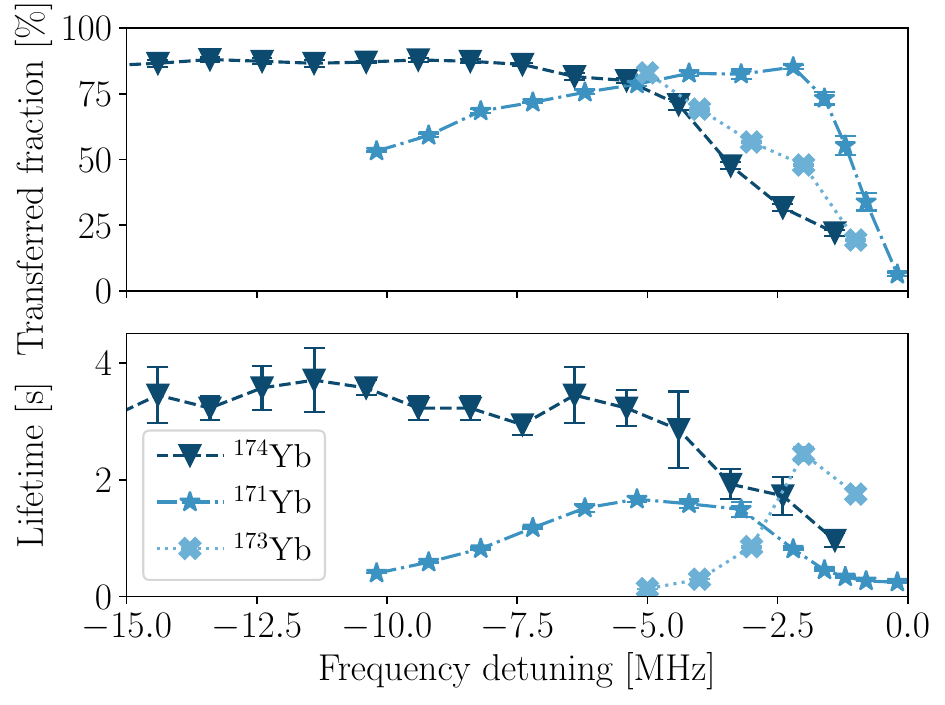"}
			\caption{\label{fig:556_pyramid_transfer_vs_detuning}Transfer efficiency and lifetime in the second MOT stage as a function of detuning for the pyramid reflector. Dashed/dotted lines are guides to the eye.}
		\end{figure}
		For our three isotopes of interest, the transfer efficiency and lifetime with the pyramid reflector is measured as a function of detuning of the $556\,$nm light. The result is shown in Fig.~\ref{fig:556_pyramid_transfer_vs_detuning}. It can be seen that for the bosonic $^{174}$Yb, the second MOT stage can be operated in a very broad frequency range with a transfer efficiency above $85\,\%$ and a lifetime of more than $3.5\,$s. For $^{171}$Yb, the frequency range of useful operation is much narrower and the maximally achievable lifetime is $1.6\,$s. However, a transfer efficiency above $80\,\%$ is attainable. While a comparably high transfer efficiency was also measured for $^{173}$Yb, this was only at frequencies where the lifetime decreased drastically. At the optimal frequency regarding the lifetime, the transfer efficiency is only close to $50\,\%$. This less efficient second-stage MOT operation could be connected to the non-optimal operation of the first MOT stage, where the atoms might not be cooled sufficiently for an efficient transfer.
		
		\begin{figure}
		\includegraphics[width = 0.45\textwidth]{"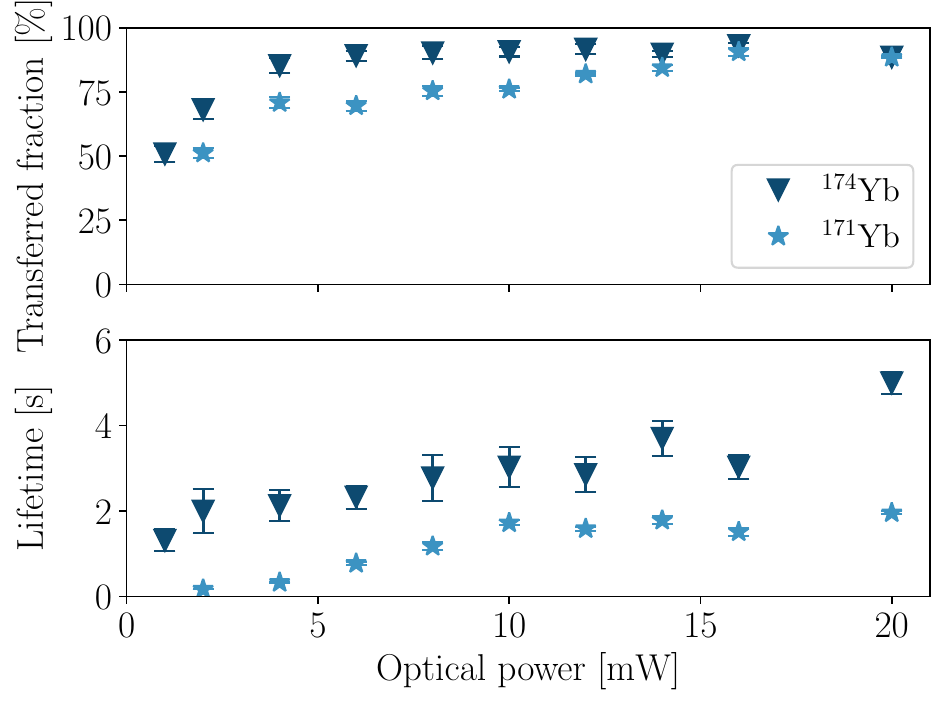"}
		\caption{\label{fig:556_pyramid_transfer_vs_power}Transfer efficiency and lifetime as a function of optical power at $556\,$nm for the pyramid reflector.}
		\end{figure}
		
		The transfer efficiency and lifetime was also measured as a function of optical power for $^{174}$Yb and $^{171}$Yb, where the detuning was set to the respective optimal values. It can be seen (Fig.~\ref{fig:556_pyramid_transfer_vs_power}) that for $^{174}$Yb, the transfer efficiency reaches its maximum just above $5\,$mW, while for $^{171}$Yb, a higher power is required for an optimal transfer. For both isotopes, the lifetime can be enhanced by increasing the power, which is due to the increased trap depth. These measurements show that a transfer to the second MOT stage is possible for all the relevant isotopes.
		
		\begin{figure}
			\includegraphics[width = 0.45\textwidth]{"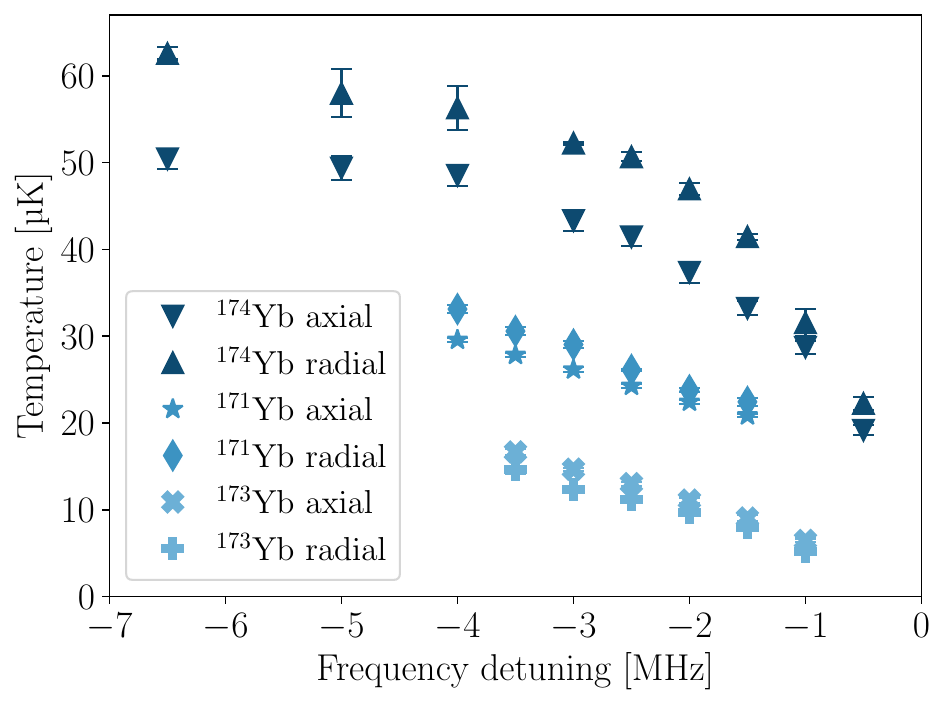"}
			\caption{\label{fig:556_pyramid_temperature}Temperature as a function of detuning in the second MOT stage for the pyramid.}
		\end{figure}
		
The temperature as a function of detuning for the three isotopes is shown in Fig.~\ref{fig:556_pyramid_temperature}. The temperature is reduced consistently for decreasing detuning.
The minimal achievable detuning was given by the decreasing transfer efficiency (see Fig.~\ref{fig:556_pyramid_transfer_vs_detuning}).
A combination of large transfer efficiency and low temperatures might be achievable by dynamically decreasing the detuning and power as well as adapting the magnetic field gradient during the transfer process~\cite{mcgrew_ytterbium_2020, pizzocaro_absolute_2017, kobayashi_uncertainty_2018a}.

For $^{174}$Yb and $^{171}$Yb, minimal temperatures around $20\,$µK were measured, which is comparable to other Yb experiments \cite{dorscher_creation_2013, kuwamoto_magnetooptical_1999}. For same detunings, the temperature of $^{171}$Yb is below the one of $^{174}$Yb. The lowest temperature is measured for $^{173}$Yb with $6.5\,$µK. This is compatible with observations in Refs. \cite{maruyama_investigation_2003, fukuhara_degenerate_2007} and is related to a sub-Doppler cooling mechanism due to the large nuclear spin \cite{kuwamoto_magnetooptical_1999}.
These results show that the pyramid reflector is well suited for two-stage cooling and trapping of the isotopes of interest.

\subsection{Fresnel reflector results}

\begin{figure}
	\includegraphics[width = 0.45\textwidth]{"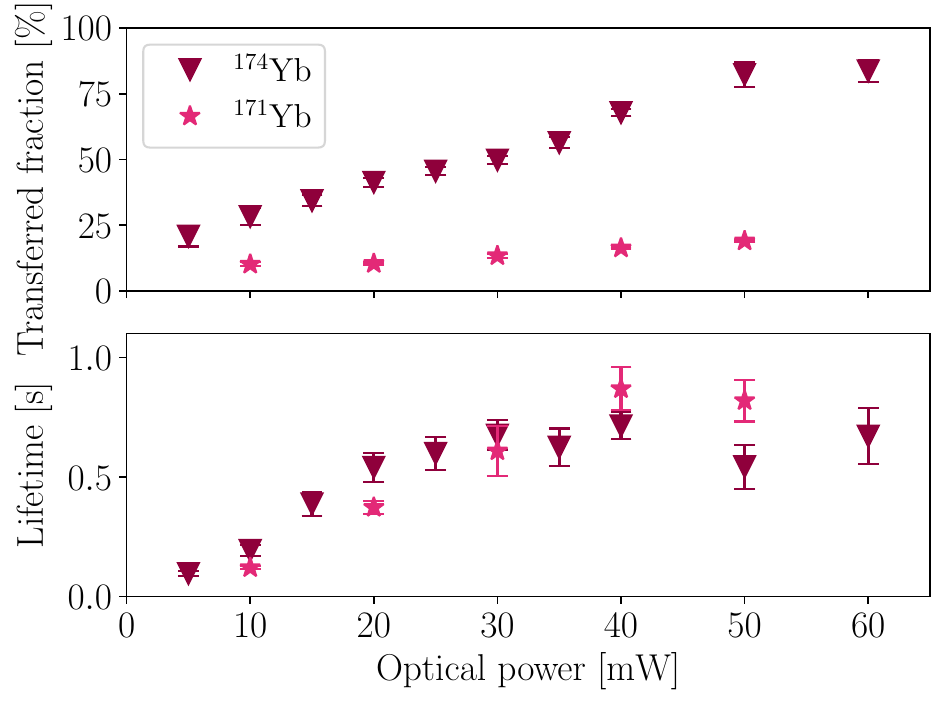"}
	\caption{\label{fig:556_fresnel_transfer_vs_power}Transfer efficiency and lifetime as a function of $556\,$nm power for the Fresnel reflector.}
\end{figure}
		
For the Fresnel reflector, the transfer efficiency and lifetime are measured as a function of $556\,$nm power. The results are shown in Fig.~\ref{fig:556_fresnel_transfer_vs_power}. It becomes apparent that a significantly higher power is required for efficient transfer than for the pyramid. For $^{174}$Yb, a transfer efficiency above $80\,\%$ is only reached  with at least $50\,$mW of optical power. Furthermore, we also observed trapping of $^{171}$Yb in the second MOT stage. The maximally measured transfer efficiency is $19\,\%$. This reduced transfer efficiency for the fermionic isotope matches the observations from the first MOT stage (Fig.~\ref{fig:399_loading_all_isotopes}) that the non-conventional beam geometry of the Fresnel reflector leads to less efficient MOT operation. For both isotopes, the lifetime in the second MOT stage is shorter than for the pyramid, all measured lifetimes are below $0.9\,$s.

		\begin{figure}
			\includegraphics[width = 0.45\textwidth]{"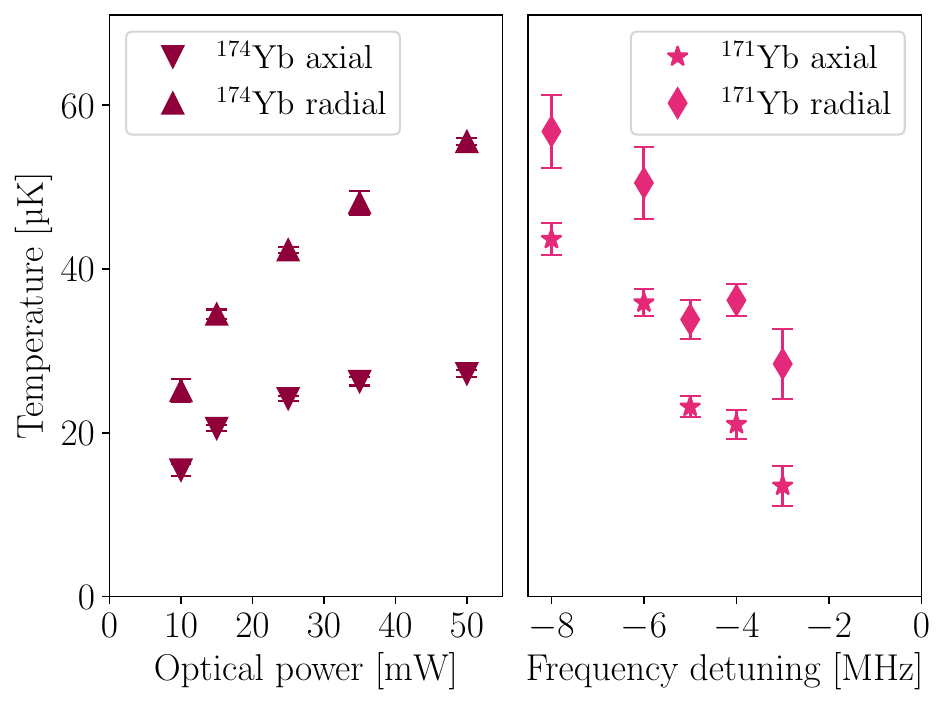"}
			\caption{\label{fig:556_fresnel_temperature}Temperature in the second MOT stage for the Fresnel reflector. For $^{174}$Yb it was found that the optical power has the largest impact on the temperature, while for $^{171}$Yb the detuning was more significant.}
		\end{figure}
		
The temperature in the second MOT stage is shown in Fig.~\ref{fig:556_fresnel_temperature}. In contrast to the pyramid reflector, for $^{174}$Yb the optical power has a significant effect on the temperature. This strong temperature dependence on the optical power can be explained by an additional heating mechanism due to spatial intensity fluctuations \cite{chaneliere_extraheating_2005} that scale with optical power. Because of the periodic mirror structure, spatial intensity fluctuations can be expected to be larger for the Fresnel reflector than for the pyramid reflector. The detuning on the other hand did not have a notable impact on the temperature as it had for the pyramid.

For $^{171}$Yb, the temperature depends more strongly on the detuning, while no effect from changing the power was observed. For both isotopes, temperatures below $30\,$µK were reached, not far from the results for the pyramid reflector. Again, the optimal parameters regarding the temperature deviate from the optimal parameters for transfer into the second MOT stage, which motivates an experimental sequence where detuning and power are simultaneously ramped down. Such a sequence could potentially enable the measurement of even lower temperatures close to the Doppler limit. These results show that second-stage cooling and trapping of the relevant isotopes is also possible in such a non-convential MOT geometry, making the Fresnel reflector a suitable candidate for the realization of compact OLCs.

	\section{Conclusion}
	We demonstrated two-stage cooling of Yb with two compact MOT setups using in-vacuum reflectors. The pyramid reflector leads to a conventional beam geometry, whereas the Fresnel reflector generates a tetrahedral MOT geometry. By using the same atomic source for loading both reflector geometries, we were able to directly compare the atom loading and transfer efficiencies. Efficient transfer into the second MOT stage was demonstrated with the pyramid reflector for $^{174}$Yb and $^{171}$Yb, with cooling into the range of $20\,$µK. Similarly high transfer efficiencies at higher optical powers were measured with the Fresnel reflector for $^{174}$Yb. Furthermore, we demonstrated second-stage cooling of the fermionic $^{171}$Yb with the Fresnel reflector. While we observe similarly low temperatures as for the bosonic isotope, the transfer efficiency for the fermion is significantly lower. This agrees with our observations in the first MOT stage, that the trapping of fermions favors conventional (pyramid) over non-conventional (Fresnel) MOT geometries.
	However, the demonstrated trapping and two-stage cooling of a fermionic alkaline-earth-like isotope in a non-conventional geometry now opens the path to transportable OLCs with a higher degree of miniaturization.
	
	\begin{acknowledgments}
	We thank the VLBAI team for fruitful discussions about the atomic source and L. Günster for the setup and characterization of the 2D-MOT. We specially thank Saskia Bondza for helpful discussions about the Fresnel MOTs trapping dynamics and for sharing experience in trapping strontium with it.
    We acknowledge funding by the Deutsche Forschungsgemeinschaft (DFG, German Research Foundation) under Germany’s Excellence Strategy – EXC-2123 QuantumFrontiers – 390837967. 
    We further acknowledge funding from the joint project ``Innovative Vacuum Technology for Quantum Sensors'' (InnoVaQ) funded by the German Federal Ministry of Education and Research (BMBF) as part of the funding program ``quantum technologies – from basic research to market''. (Contract number: 13N15915). 
	\end{acknowledgments}

\bibliography{complete_bibliography}
		
\end{document}